\documentclass[journal,onecolumn,12pt]{IEEEtran} 
\usepackage[]{times} 
\usepackage{epsfig} 
\usepackage{subfigure}
\usepackage{amsmath} 
\usepackage{amssymb} %\usepackage{amsthm} 
\usepackage{graphicx} 
\usepackage{subfigure}
\usepackage{multirow} 
\renewcommand{\baselinestretch}{2} % For 1.5
\newcommand{\bega}{\begin{eqnarray}}
\newcommand{\ega}{\end{eqnarray}}
\newcommand{\bb}{\begin{equation}}
\newcommand{\ee}{\end{equation}}
\newtheorem{defn} {Definition}
\newtheorem{te}{Theorem}
\newtheorem{lema}{Lemma}
\newtheorem{cor}{Corollary}
\newtheorem{ex}{Example}

\begin{document}
\title{Error Correction Capability of Column-Weight-Three LDPC Codes}

\author{Shashi~Kiran~Chilappagari,~\IEEEmembership{Student~Member,~IEEE,}
        and~Bane~Vasic,~\IEEEmembership{Senior Member,~IEEE}% <-this % stops a space
\thanks{Manuscript received \today. This work is funded by NSF under Grant CCF-0634969, ITR-0325979 and INSIC-EHDR program.}
\thanks{S. K. Chilappagari and B. Vasic are with the Department of Electrical and Computer Engineering, University of Arizona, Tucson, AZ, 85721 USA
(e-mails: shashic@ece.arizona.edu, vasic@ece.arizona.edu).}}% <-this % stops a space
% <-this % stops a space
%\markboth{Submitted to  IEEE Transactions on Information Theory}{Submitted to  IEEE Transactions on Information Theory}
%\markboth{IEEE Transactions On Information Theory }{Chilappagari \MakeLowercase{\textit{et al.}}: Expander Graph Arguments for Unreliable Memories}

\maketitle
\thispagestyle{empty}
\begin{abstract}
In this paper, we investigate the error correction capability of column-weight-three LDPC codes when decoded using the Gallager A algorithm. We prove that the necessary condition for a code to correct $k \geq 5$ errors is to avoid cycles of length up to $2k$ in its Tanner graph. As a consequence of this result, we show that given any $\alpha>0, \exists N $ such that $\forall n>N$, no code in the ensemble of column-weight-three codes can correct all $\alpha n$ or fewer errors. We extend these results to the bit flipping algorithm.
\end{abstract}

\begin{center} \textbf{\small Index Terms}
\end{center}

{\small Low-density parity-check codes, Gallager A algorithm, trapping sets, error correction capability}

\section{Introduction}
Gallager in \cite{gallager} showed that for $\gamma \geq 3$ and $\rho >\gamma$, there exist $(n,\gamma,\rho)$ regular low-density parity-check (LDPC) codes for which the bit error probability tends to zero asymptotically whenever we operate below the threshold. Richardson and Urbanke in \cite{richardsonurbanke} derived the capacity of LDPC codes for various message passing algorithms and described density evolution, a deterministic algorithm to compute thresholds. Zyablov and Pinsker \cite{zyablov} analyzed LDPC codes under a simpler decoding algorithm known as the bit flipping algorithm and showed that almost all the codes in the regular ensemble with $\gamma \geq 5$ can correct a constant fraction of worst case errors. Sipser and Spielman in \cite{spielman} used expander graph arguments to analyze bit flipping algorithm. Burshtein and Miller in \cite{burshtein} applied expander based arguments to show that message passing algorithms can also correct a fixed fraction of worst case errors when the degree of each variable node is at least five. Feldman \textit{et al.} \cite{feldman} showed that linear programming decoder \cite{feldman2} is also capable of correcting a fraction of errors. Recently, Burshtein in \cite{burshteinisitpaper} showed that regular codes with variable nodes of degree four  are capable of correcting a linear number of errors under bit flipping algorithm. He also showed tremendous improvement in the fraction of correctable errors when the variable node degree  is at least five.

In this paper, we consider the error correction capability of the ensemble $\mathcal{C}^n(3,\rho>3)$ of $(3,\rho)$ regular LDPC codes as defined in \cite{richardsonurbanke} when decoded using the Gallager A algorithm. We analyze decoding failures using the notion of trapping sets and prove that a code with girth $g \geq 10$ cannot correct all $g/2$ or fewer errors. Using this result, we prove that for any $\alpha >0$, for sufficiently large block length $n$, no code in the $\mathcal{C}^n(3,\rho)$ ensemble can correct $\alpha$ fraction of errors. This result settles the problem of error correction capability of column-weight-three codes. The rest of the paper is organized as follows. In Section \ref{section2} we establish the notation and describe the Gallager A algorithm. We then characterize the failures of the Gallager A decoder with the help of fixed points. We also introduce the notions of trapping sets, failure sets and critical number. In Section \ref{section3} we investigate the relation between error correction capacity and girth of the code. We extend the results to bit flipping algorithm in Section \ref{section4} and conclude in Section \ref{section5}.

\section{Decoding Algorithms and Trapping Sets}\label{section2}

\subsection{Graphical Representations of LDPC Codes}
LDPC codes \cite{gallager} are a class of linear block codes which can be defined by sparse bipartite graphs \cite{shokrollahi}. Let $\cal{G}$ be a bipartite graph with two sets of nodes: $n$ variable nodes and $m$ check nodes. The check nodes (variable nodes) connected to a variable node (check node) are referred to as its neighbors. The degree of a node is the number of its neighbors. This graph defines a linear block code $\mathcal{C}$ of length $n$ and dimension at least $n-m$ in the following way: The $n$ variable nodes are associated to the $n$ coordinates of codewords. A vector $\mathbf{v}=(v_1,v_2,\ldots,v_n)$ is a codeword if and only if for each check node, the sum  of its neighbors is zero. Such a graphical representation of an LDPC code is called the Tanner graph \cite{tanner} of the code. The adjacency matrix of $\cal{G}$ gives $H$, a parity check matrix of $\cal{C}$. An $(n,\gamma,\rho)$ regular LDPC code has a Tanner graph with $n$ variable nodes each of degree  $\gamma$ (column weight) and $n\gamma/ \rho$ check nodes each of degree  $\rho$ (row weight). This code has length $n$ and rate $r \geq 1-\gamma/\rho$ \cite{shokrollahi}. In the rest of the paper we consider codes in the $(3,\rho)$, $\rho>3$, regular LDPC code ensemble. Note that the column weight and row weight are also referred to as left degree and right degree in literature. It should also be noted that the Tanner graph is not uniquely defined by the code and when we say the Tanner graph of an LDPC code, we only mean one possible graphical representation. The girth $g$ is the length of the shortest cycle in $\cal{G}$. In this paper, $\bullet$ represents a variable node, $\square$ represents an even degree check node and $\blacksquare$ represents an odd degree check node. 

\subsection{Hard Decision Decoding Algorithms}
Gallager in \cite{gallager} proposed two simple binary message passing algorithms for decoding over the binary symmetric channel (BSC); Gallager A and Gallager B. See \cite{shokrollahi} for a detailed description of Gallager B algorithm. For column-weight-three codes, which are the main focus of this paper, these two algorithms are the same. Every round of message passing (iteration) starts with sending messages from variable nodes (first half of the iteration) and ends by sending messages from check nodes to variable nodes (second half of the iteration). Let $\mathbf{r}$, a binary $n$-tuple be the input to the decoder. Let $\omega_j(v,c)$ denote the message passed by a variable node $v$ to its neighboring check node $c$ in $j^{th}$ iteration and $\varpi_j(c,v)$ denote the message passed by a check node $c$ to its neighboring variable node $v$. Additionally, let $\omega_j(v,\colon)$ denote the set of all messages from $v$, $\omega_j(v,\colon \backslash c)$ denote the set of all messages from $v$ except to $c$, $\omega_j(\colon,c)$ denote the set of all messages to $c$. $\omega_j(\colon \backslash v,c),\varpi_j(c,\colon),\varpi_j(c,\colon \backslash v), \varpi_j(\colon,v)$ and $\varpi_j(\colon \backslash c,v)$ are defined similarly. The Gallager A algorithm can be defined as follows.
\begin{eqnarray}
\omega_1(v,c)&=&\mathbf{r}(v) \nonumber \\ 
\omega_j(v,c) &=& \left \{ \begin{array}{cl}
												1, & \mbox{if } \varpi_{j-1}(\colon \backslash c,v)=1\\
											0, & \mbox{if } \varpi_{j-1}(\colon \backslash c,v)=0\\
											\mathbf{r}(v), & \mbox{otherwise}  \end{array}\right. \nonumber \\
\varpi_j(c,v)&=& \left(\sum \omega_j(\colon \backslash v,c)\right) \mbox{mod } 2 \nonumber
\end{eqnarray}

At the end of each iteration, an estimate of each variable node is made based on the incoming messages and possibly the received value. The decoded word at the end of $j^{th}$ iteration is denoted as $\mathbf{r}^{(j)}$. The decoder is run  until a valid codeword is found or a maximum number of iterations $M$ is reached, whichever is earlier. The output of the decoder is either a codeword or $\mathbf{r}^{(M)}$.

\textit{A Note on the Decision Rule:} Different rules to estimate a variable node after each iteration are possible and it is likely that changing the rule after certain iterations may be beneficial. However, the analysis of various scenarios is beyond the scope of this paper. For column-weight-three codes only two rules are possible.
\begin{itemize}
\item Decision Rule A: if all incoming messages to a variable node from neighboring checks are equal, set the variable node to that value; else set it to received value
\item Decision Rule B: set the value of a variable node to the majority of the incoming messages; majority always exists since the column-weight is three 
\end{itemize}
We adopt Decision Rule A throughout this paper.
\subsection{Trapping Sets of Gallager A Algorithm}
We now characterize  failures of the Gallager A decoder using fixed points and trapping sets \cite{rich}. Consider an LDPC code of length $n$ and let $\mathbf{x}$ be the binary vector which is the input to the Gallager A decoder. Let $\mathcal{S}(\mathbf{x})$ be the support of $\mathbf{x}$. The support of $\mathbf{x}$ is defined as the set of all positions $i$ where $\mathbf{x}(i) \neq 0$. Without loss of generality, we assume that the all zero codeword is sent over BSC and that the input to the decoder is the error vector. Hence, throughout this paper a message of $1$ is alternatively referred to as an incorrect message, a received value of $1$ is referred to as an initial error. 
\begin{defn}\cite{rich}
A decoder failure is said to have occurred if the output of the decoder is not equal to the transmitted codeword.
\end{defn}

\begin{defn}
$\mathbf{x}$ is called a \textit{fixed point} if 
\begin{eqnarray}
\omega_j(v,c)=\mathbf{x}(v),~~\forall j > 0 \nonumber
\end{eqnarray}
\end{defn}

That is, the message passed from variable nodes to check nodes along the edges are the same in every iteration. Since the outgoing messages from variable nodes are same in every iteration, it follows that the incoming messages from check nodes to variable nodes are also same in every iteration and so is the estimate of a variable after each iteration. In fact, the estimate after each iteration coincides with the received value. It is clear from above definition that if the input to the decoder is a fixed point, then the output of the decoder is the same fixed point. 

\begin{defn}\cite{chilappagarione}
Let $\mathbf{x}$ be a fixed point. Then $\mathcal{S}(\mathbf{x})$ is known as a trapping set. A $(V,C)$ trapping set $\cal{T}$ is a set of $V$ variable nodes whose induced subgraph has $C$ odd degree checks. 
\end{defn}

\begin{te}\label{thm1}
Let $\mathcal{C}$ be a code in the ensemble of $(3,\rho)$ regular LDPC codes. Let $\cal{T}$ be a set consisting of $V$ variable nodes with induced subgraph $\cal{I}$. Let the checks in $\cal{I}$ be partitioned into two disjoint subsets; $\cal{O}$ consisting of checks with odd degree and $\cal{E}$ consisting of checks with even degree. Let $|\mathcal{O}|=C$ and $|\mathcal{E}|=S$.  $\cal{T}$ is a trapping set iff : (a) Every variable node in $\cal{I}$ is connected to at least two  checks in $\cal{E}$ and (b) No two checks of $\cal{O}$ are connected to a variable node outside $\cal{I}$.
\end{te}

\begin{proof} See Appendix \ref{appendix1}
\end{proof}

We note that Theorem \ref{thm1} is a consequence of Fact 3 in \cite{rich}. We also remark that Theorem \ref{thm1} can be extended to higher column weight codes but in this paper we restrict our attention to column-weight-three codes. 

If the variable nodes corresponding to a trapping set are in error, then a decoder failure occurs. However, not all variable nodes corresponding to  trapping set need to be in error for a decoder failure to occur.

\begin{defn}\cite{chilappagarione} The minimal number of variable nodes that have to be initially in error for the decoder to end up in the trapping set $\cal{T}$ will be referred to as {\it critical number} for that trapping set.\end{defn}

\begin{figure*}[htb]
\centering
\subfigure[] % caption for subfigure a
{
    \label{sixcycle}

\includegraphics[width=0.3\textwidth]{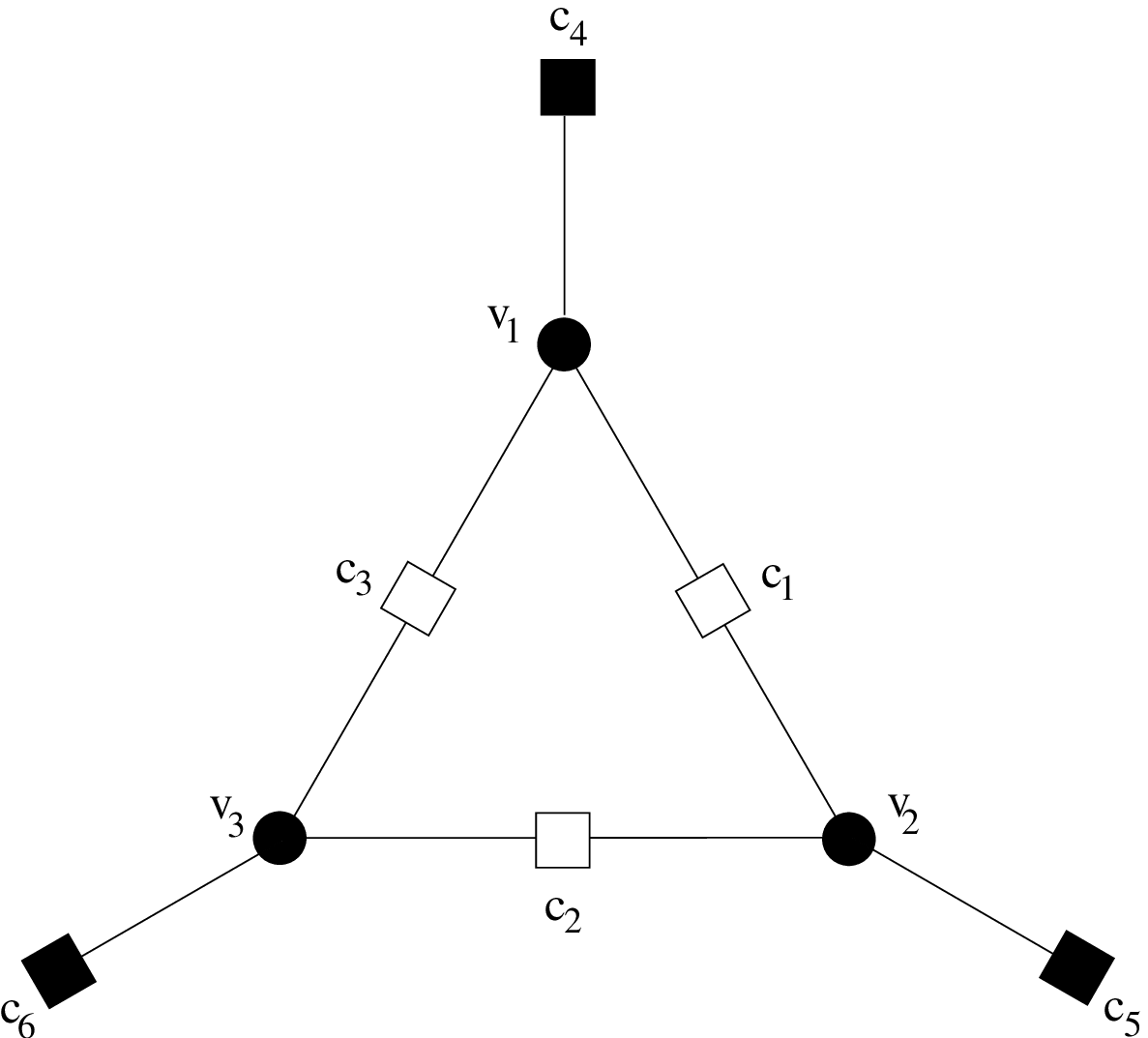}
}
\hspace{0.1\textwidth}
\subfigure[] % caption for subfigure a
{
    \label{42trappingset}

\includegraphics[width=0.3\textwidth]{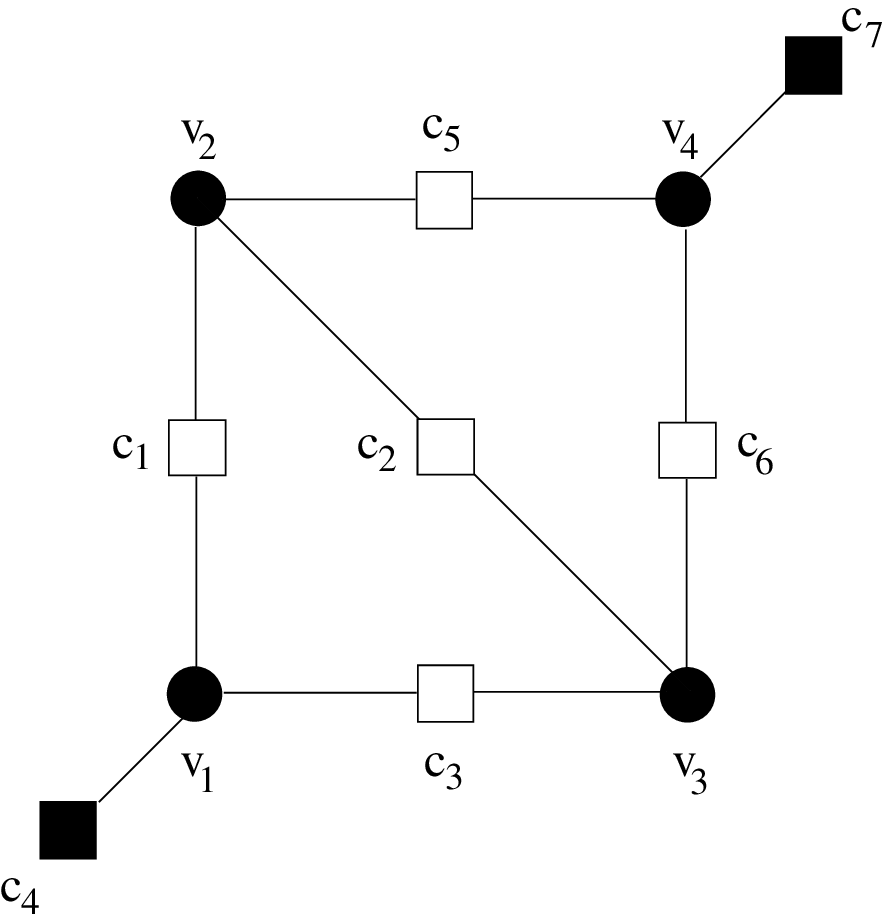}
}
\caption{Examples of trapping sets \subref{sixcycle} a $(3,3)$ trapping set \subref{42trappingset} a $(4,2)$ trapping set} \label{3trappingsets}
\end{figure*}

\begin{defn}A set of variable nodes which if in error lead to a decoding failure is known as a \textit{failure set}.\end{defn}
\textit{Remarks}
\begin{enumerate}
\item To ``end up'' in a trapping set $\cal{T}$ means that, after a possible finite  number of iterations, the decoder will be in error, on at least one variable node from $\cal{T},$ at every iteration \cite{rich}. 
\item The notion of a failure set is more fundamental than a trapping set. However, from the definition, we cannot derive necessary and sufficient conditions for a set of variable nodes to form a failure set.
\item A trapping set is a failure set. Subsets of trapping sets can be failure sets. More specifically, for a trapping set of size $V$, there exists at least one subset of size equal to the critical number which is a failure set.
\item The critical number of a trapping set is not fixed. It depends on the outside connections of checks in $\cal{E}$. However, the maximum value of critical number of a $(V,C)$ trapping set is $V$.
\end{enumerate}
\begin{ex}
Fig.\ref{sixcycle} shows a subgraph induced by a set of three variable nodes $\{v_1,v_2,v_3\}$ . If no two odd degree check nodes from $\{c_4,c_5,c_6\}$ are connected to a variable outside the subgraph, then by Theorem \ref{thm1}, Fig.\ref{sixcycle} is a $(3,3)$ trapping set. On the other hand, if two odd degree checks, say $c_5$ and $c_6$, are connected to another variable node, say $v_4$, the subgraph resembles Fig. \ref{42trappingset}. Assuming no other connections, Fig.\ref{42trappingset} is a $(4,2)$ trapping set. We make the following observations:
\begin{enumerate}
\item The three variable nodes in a $(3,3)$ trapping set form a six cycle. However, not all six cycles are $(3,3)$ trapping sets. Apart from the subgraph induced by variable nodes, the outside connections should be known to determine whether a given subgraph is a trapping set or not. The $(4,2)$ trapping set in Fig.\ref{42trappingset} illustrates this point. 
\item The critical number of a $(3,3)$ trapping set is three. There exist $(4,2)$ trapping sets with critical number three and it is highly unlikely that a $(4,2)$ trapping set does not contain a failure set of size three. However, we can show by a counterexample that this is indeed possible.
\item A $(V,C)$ trapping set is not unique i.e., two trapping sets with same $V$ and $C$ can have different underlying topological structures (induced subgraphs). So, when we talk of a trapping set, we refer to a specific topological structure. In this paper, the induced subgraph is assumed to be known from the context.
\item To avoid a trapping set in a code, it is sufficient to avoid topological structures isomorphic to the subgraph induced by the trapping set. For example to avoid $(3,3)$ trapping sets of Fig.\ref{sixcycle}, it is sufficient to avoid six cycles. It is possible that a code has six cycles but no $(3,3)$ trapping sets. In this case all six cycles are part of $(4,2)$ or other trapping sets.  
\end{enumerate}
\end{ex}

\section{Error Correction Capability and Girth of the Code}\label{section3}
Burshtein and Miller in \cite{burshtein} applied expander based arguments to message passing algorithms. They analyzed ensembles of irregular graphs and showed that if the degree of each variable node is at least five, then message passing algorithms can correct a fraction of errors. Codes with column weight three and four cannot achieve the expansion required for these arguments. Recently, Burshtein in \cite{burshteinisitpaper} developed a new technique to investigate the error correction capability of regular LDPC codes and showed that at sufficiently large block lengths, almost all codes with column weight four are also capable of correcting a fraction of errors under bit flipping algorithm. For column-weight-three codes he notes that such a result cannot be proved. This is because a non negligible fraction of codes have parallel edges in their Tanner graphs and such codes cannot correct a single worst case error. 

In this paper we prove a stronger result by showing that for any given $\alpha > 0$, at sufficiently large block lengths $n$, no code in the $\mathcal{C}^n(3,\rho)$ ensemble can correct all $\alpha n$ or fewer errors under Gallager A algorithm and show that this holds for the bit flipping algorithm also.  

\begin{lema}\cite{burshteinisitpaper} \label{lemma1}
A code whose Tanner graph has parallel edges cannot correct a single worst case error.
\end{lema}
\begin{proof}
See \cite{burshteinisitpaper}. The proof is for bit flipping algorithm, but also applies to Gallager A algorithm.
\end{proof}
\begin{lema}\label{lemma2}
Let $\cal{C}$ be an $(n,3,\rho)$ regular LDPC code with girth $g=4$. Then $\cal{C}$ has at least one failure set of size two or three. 
\end{lema}
\begin{proof}
See Appendix \ref{appendix2}.
\end{proof}
\begin{lema}\label{lemma3}
Let $\cal{C}$ be an $(n,3,\rho)$ regular LDPC code with girth $g=6$. Then $\cal{C}$ has least one failure set of size three or four.
\end{lema}
\begin{proof}
Since $g=6$, there is at least one six cycle. Without loss of generality, we assume that $\{v_1,v_2,v_3\}$ together with the three even degree checks  $\{c_1,c_2,c_3\}$ and the three odd degree checks $\{c_4,c_5,c_6\}$ form a six cycle as in Fig.\ref{sixcycle}. If no two checks from $\{c_4,c_5,c_6\}$ are connected to a variable node, then $\{v_1,v_2,v_3\}$ is a $(3,3)$ trapping set and hence a failure set of size three. On the contrary, assume that $\{v_1,v_2,v_3\}$ do not form a $(3,3)$ trapping set. Then there exists $v_4$ which is connected to at least two checks from $\{c_4,c_5,c_6\}$. If $v_4$ is connected to all the three checks, $\{v_1,v_2,v_3,v_4\}$ is a codeword of weight four and it is easy to see that $\{v_1,v_2,v_3\}$ is a failure set. Now assume that $v_4$ is connected to only two checks from $\{c_4,c_5,c_6\}$. Without loss of generality, let the two checks be $c_5$ and $c_6$. Let the third check connected to $v_4$ be $c_7$ as shown in Fig.\ref{42trappingset}. If $c_4$ and $c_7$ are not connected to a common variable node then $\{v_1,v_2,v_3,v_4\}$ is a $(4,2)$ trapping set and hence a failure set of size four. If $c_4$ and $c_7$ are connected to say $v_5$, we have two possibilities: (a) The third check is $c_8$ and (b) The third check of $v_5$ is $c_2$ (the third check cannot be $c_1$ or $c_3$ as this would introduce a four cycle). We claim that in both cases $\{v_1,v_2,v_3,v_4\}$ is a failure set. The two cases are discussed below.

Case (a): Let $S(\mathbf{r})=\{v_1,v_2,v_3,v_4\}$. 
\begin{eqnarray}
\omega_1(v,:)&=&\left\{\begin{array}{cl}
1, & v \in \{v_1,v_2,v_3,v_4\} \\
0, & \mbox{otherwise}
\end{array} \right. \nonumber 
\end{eqnarray}
The messages in the second half of first iteration are, 
\begin{eqnarray}
\varpi_1(c_1,v)&=&\left\{\begin{array}{cl}
1, & v \in \{v_1,v_2\} \\
0, & \mbox{otherwise}
\end{array} \right. \nonumber 
\end{eqnarray}
Similar equations hold for $c_2,c_3,c_5,c_6$. For $c_4$ we have
\begin{eqnarray}
\varpi_1(c_4,v)&=&\left\{\begin{array}{cl}
0, & v = v_1  \\
1, & \mbox{otherwise}
\end{array} \right. \nonumber 
\end{eqnarray}
Similar equations hold for $c_7$. At the end of first iteration, we note that $v_2$ and $v_3$ receive all incorrect messages, $v_1,v_4$ and $v_5$ receive two incorrect messages and all other variable nodes receive at most one incorrect message. We therefore have $\mathbf{r}^{(1)}=\mathbf{r}$ and $S(\mathbf{r}^{(1)})=\{v_1,v_2,v_3,v_4\}$. The messages sent by variable nodes in the second iteration are, 
\begin{eqnarray}
\omega_2(v,:) &=& 1,  v \in \{v_1,v_2,v_3,v_4\} \nonumber \\
\omega_2(v_5,c_8)&=&1, \nonumber \\
\omega_2(v_5,\{c_4,c_7\})&=&0, \nonumber \\
\omega_2(v,:)&=&0, v \in \{v_1,\ldots,v_n\} \setminus \{v_1,v_2,v_3,v_4,v_5\}. \nonumber
\end{eqnarray}
The messages passed in second half of second iteration are same as in second half of first iteration, except that $\varpi(c_8,\colon \backslash v_5)=1$. At the end of second iteration, we note that $v_2$ and $v_3$ receive all incorrect messages, $v_1,v_4$ and $v_5$ receive two incorrect messages and all other variable nodes receive at most one incorrect message. The situation is same as at the end of first iteration. The algorithm runs for \textit{M} iterations and the decoder outputs $\mathbf{r}^{(M)}=\mathbf{r}$ which implies that $\{v_1,v_2,v_3,v_4\}$ is a failure set.

Case (b): The proof is along the same lines as for Case (a). The messages for first iteration are the same. The messages in the first half of second iteration are,
\begin{eqnarray}
\omega_2(v,:) &=& 1,  v \in \{v_1,v_2,v_3,v_4\} \nonumber \\
\omega_2(v_5,c_2)&=&1, \nonumber \\
\omega_2(v_5,\{c_4,c_7\})&=&0, \nonumber \\
\omega_2(v,:)&=&0, v \in \{v_1,\ldots,v_n\} \setminus \{v_1,v_2,v_3,v_4,v_5\}. \nonumber
\end{eqnarray}
The messages passed in second half of second iteration are same as in second half of first iteration, except that $\varpi(c_2,\colon \backslash \{v_2,v_3,v_5\})=1$ and $\varpi(c_2,\{v_2,v_3,v_5\})=0$ . At the end of second iteration, $v_1,v_2,v_3,v_4$ and $v_5$ receive two incorrect messages and all other variable nodes receive at most one incorrect message and hence $\mathbf{r}^{(2)}=\mathbf{r}$. The messages passed in first half of third iteration (and therefore subsequent iterations) are same as the messages passed in first half of second iteration.  The algorithm runs for \textit{M} iterations and the decoder outputs $\mathbf{r}^{(M)}=\mathbf{r}$ which implies that $\{v_1,v_2,v_3,v_4\}$ is a failure set.
\end{proof}

\begin{lema}\label{lemma4}
Let $\cal{C}$ be an $(n,3,\rho)$ regular LDPC code with girth $g=8$. Then $\cal{C}$ has at least one failure set of size four or five.
\end{lema}
\begin{proof}
See Appendix \ref{appendix2}.
\end{proof}
\textit{Remark:} It might be possible that Lemmas \ref{lemma2}--\ref{lemma4} can be made stronger by further analysis, i.e., it might be possible to show that a code with girth four has a failure set of size two, a code with girth six has failure set of size three and a code with girth eight has a failure set of size four. However, these weaker lemmas are sufficient to establish the main theorem.

\begin{figure*}[htb]
\centering
\includegraphics[width=0.5\textwidth]{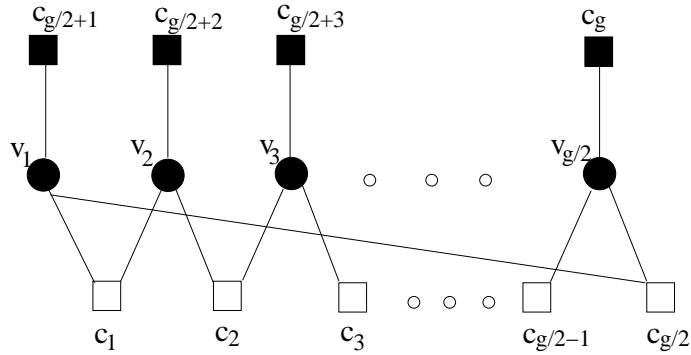}
\caption{Illustration of a cycle of length $g$} \label{mincycle}
\end{figure*}

\begin{lema}\label{lemma5}
Let $\cal{C}$ be an $(n,3,\rho)$ regular LDPC code with girth $g \geq 10$. Then the set of variable nodes $\{v_1,v_2,\ldots,v_{g/2}\}$ involved in the shortest cycle is a trapping set of size $g/2$. \end{lema}
\begin{proof}
Since $\cal{C}$ has girth $g$, there is at least one cycle of length $g$. Without loss of generality, assume that $\{v_1,v_2,\ldots,v_{g/2}\}$ form a cycle of minimum length as shown in Fig.\ref{mincycle}. Let the even degree checks be $\mathcal{E}=\{c_1,c_2,\ldots,c_{g/2}\}$ and the odd degree checks be $\mathcal{O}=\{c_{g/2+1},c_{g/2+2},\ldots,c_{g}\}$. Note that each variable node is connected to two checks from $\cal{E}$ and one check from $\cal{O}$ and $c_{g/2+i}$ is connected to $v_i$. We claim that no two checks from $\cal{O}$ can be connected to a common variable node. 

The proof is by contradiction. Assume $c_i$ and $c_j$ ($g/2+1 \leq i<j \leq g)$ are connected to a variable node $v_{ij}$. Then $\{v_i,\ldots,v_j,v_{ij}\}$ form a cycle of length $2(j-i+2)$ and $\{v_j,\ldots\,v_{g/2},v_1,\ldots,v_i,v_{ij}\}$ form a cycle of length $2(g/2-j+i+2)$. Since $g \geq 10$, 
\[\min(2(j-i+2),2(g/2-j+i+2)) < g.
\] 
This implies that there is a cycle of length less than $g$, which is a contradiction as the girth of the graph is $g$. 

By Theorem \ref{thm1}, $\{v_1,v_2,\ldots,v_{g/2}\}$ is a trapping set.
\end{proof}

\begin{cor}\label{corollary1}
For a code to correct all $k \geq 5$ or fewer errors, it is necessary to avoid all cycles up to length $2k$.
\end{cor}
We now state and prove the main theorem.
\begin{te}\label{thm2}
Consider the standard $(3,\rho)$ regular LDPC code ensemble. Let $\alpha > 0$. Let $N$ be the smallest integer satisfying
\begin{eqnarray}
\alpha N &>& 2 \left(\frac{\log{N}}{\log{(2(\rho-1))}}+1 \right)   \nonumber \\
\alpha N &\geq& 5. \nonumber 
\end{eqnarray}
Then, for $n > N$, no code in the $\mathcal{C}^n$$(3,\rho)$ ensemble can correct all $\alpha n$ or fewer errors.
\end{te}
\begin{proof}
First observe that for any $n>N$, we have
\begin{eqnarray}\label{girtheq1}
\alpha n > 2 \left (\frac{\log{n}}{\log{(2(\rho-1))}}+1\right).
\end{eqnarray}

From [Theorem C.1 \cite{gallager}] and [Lemma C.1 \cite{gallager}], we have the girth $g$ of any code in $\mathcal{C}^n(3,\rho)$ is bounded by
\begin{eqnarray}\label{girtheq2}
g \leq 4 \left( \frac{\log{n}}{\log{(2(\rho-1))}}+1 \right )
\end{eqnarray}
For $n>N$, Equations (\ref{girtheq1}) and (\ref{girtheq2}) imply that for any code in the $\mathcal{C}^n(3,\rho)$ ensemble, the girth is bounded by
\[
g < 2\alpha n.
\]  
The result now follows from Corollary \ref{corollary1}.
\end{proof}
\section{Extension to the Bit Flipping Algorithm} \label{section4}
The bit flipping algorithm does not belong to the class of message passing algorithms. However, the definitions from Section \ref{section2} and the results from Section \ref{section3} can be generalized to the parallel bit flipping algorithm \cite{spielman}. Without loss of generality we assume that the all zero codeword is sent. We begin with a few definitions.
\begin{defn}\cite{spielman}
A variable node is said to be corrupt if it is different from its original sent value. In our case, a variable node is corrupt if it is $1$. A check node is said to be satisfied if it is connected to even number of corrupt variables and unsatisfied otherwise.
\end{defn}
\begin{defn}
Let $\mathbf{r}$ be the input to the parallel bit flipping decoder. $S(\mathbf{r})$ is a trapping set for bit flipping algorithm if the set of corrupt variables after every iteration is $S(\mathbf{r})$.
\end{defn}
\begin{te}\label{thm3}
Let $\mathcal{T}$ be a set of variable nodes satisfying the conditions of Theorem \ref{thm1}. Then $\mathcal{T}$ is a trapping set for the bit flipping algorithm.
\end{te}
\begin{proof}
Let $S(\mathbf{r})=\mathcal{T}$. Then $\mathcal{T}$ is the set of corrupt variable nodes. Observe that a variable flips if it is connected to at least two unsatisfied checks. Since no variable is connected to two unsatisfied checks, the set of corrupt variable nodes is unchanged and by definition $\mathcal{T}$ is a trapping set.
\end{proof}
We note that Theorem \ref{thm3} is also a consequence of Fact 3 from \cite{rich}.
\begin{cor}
A trapping set for Gallager A is also a trapping set for bit flipping algorithm. 
\end{cor}
It can be shown that Lemmas \ref{lemma1}--\ref{lemma5} and Theorem \ref{thm2} also hold for the bit flipping algorithm.
\section{Conclusion}\label{section5}
In this paper we have investigated the error correction capability of column-weight-three codes under Gallager A and extended the results to bit flipping algorithm. Future work includes investigation of sufficient conditions to correct a given number of errors for column-weight-three as well as higher column weight codes.

\appendices
\section{} \label{appendix1}
\textit{Proof of Theorem \ref{thm1}:} Let $\mathbf{r}$ be the input to the decoder with $S(\mathbf{r})=\cal{T}$. Then,
\begin{eqnarray}
\omega_1(v,:)&=&\left\{\begin{array}{cl}
1, & v \in \cal{T} \\
0, & \mbox{otherwise}
\end{array} \right. \nonumber 
\end{eqnarray}
Let a check node $c_o \in \cal{O}$. Then,
\begin{eqnarray}
\varpi_1(c_o,v)&=&\left\{\begin{array}{cl}
0, & v \in \cal{T} \\
1, & \mbox{otherwise}
\end{array} \right. \nonumber 
\end{eqnarray}
Let a check node $c_e \in \cal{E}$. Then,
\begin{eqnarray}
\varpi_1(c_e,v)&=&\left\{\begin{array}{cl}
1, & v \in \cal{T} \\
0, & \mbox{otherwise}
\end{array} \right. \nonumber 
\end{eqnarray}
For any other check node $c$, $\varpi_1(c,v)=0$. By the conditions of the theorem, at the end of first iteration, any $ v \in \mathcal{T}$ receives at least two $1$'s and any $v \notin \mathcal{T}$ receives at most one $1$. So, we have
\begin{eqnarray}
\omega_2(v,:)&=&\left\{\begin{array}{cl}
1, & v \in \cal{T} \\
0, & \mbox{otherwise}
\end{array} \right. \nonumber 
\end{eqnarray}
By definition, $\mathcal{T}$ is a trapping set. 

To see that the conditions stated are necessary observe that for a variable node to send the same messages as in the first iteration, it should receive at least two messages which coincide with the received value.

\hfill $\blacksquare$
\section{}\label{appendix2}
\textit{Proof of Lemma \ref{lemma2}:} Let $\{v_1, v_2\}$ be the variable nodes that form a four cycle with even degree checks $\{c_1,c_2\}$ and odd degree checks $\{c_3,c_4\}$. If $c_3$ and $c_4$ are not connected to a common variable node, then $\{v_1,v_2\}$ is a $(2,2)$ trapping set and hence a failure set of size two.  Now assume that $c_3$ and $c_4$ are connected to a common variable node $v_3$. Then, $\{v_1,v_2,v_3\}$ is a $(3,1)$ trapping set and therefore a failure set of size three. \hfill $\blacksquare$

\begin{figure*}[htb]
\centering
\subfigure[] % caption for subfigure a
{
    \label{44trappingset}

\includegraphics[width=0.3\textwidth]{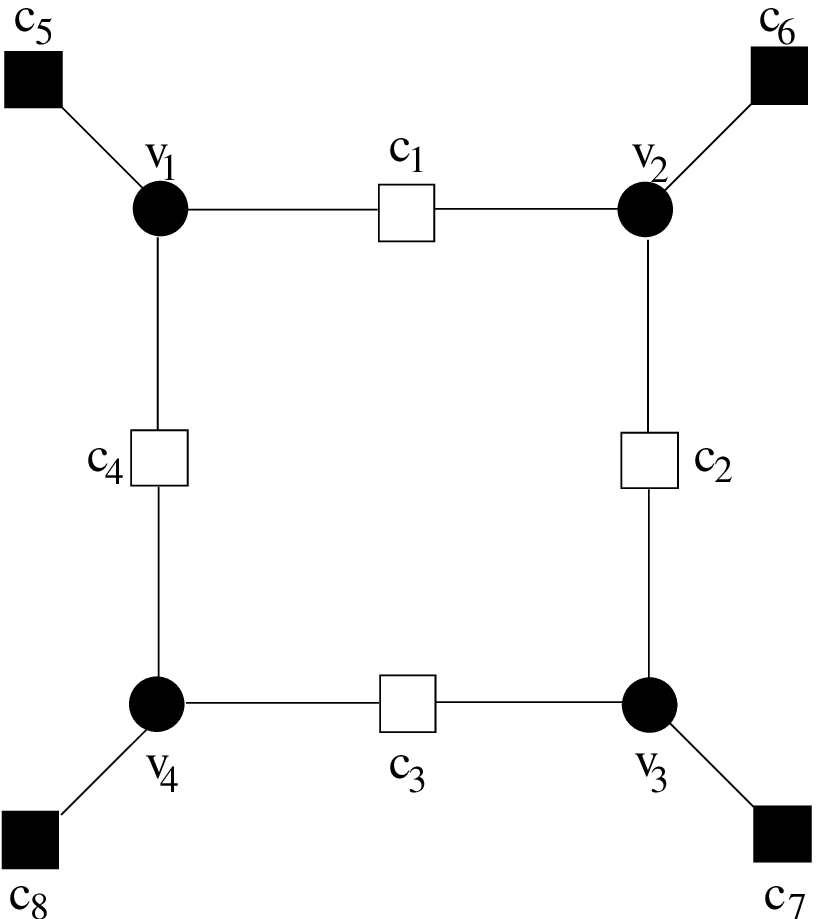}
}
\hspace{0.1\textwidth}
\subfigure[] % caption for subfigure a
{
    \label{53trappingset}

\includegraphics[width=0.3\textwidth]{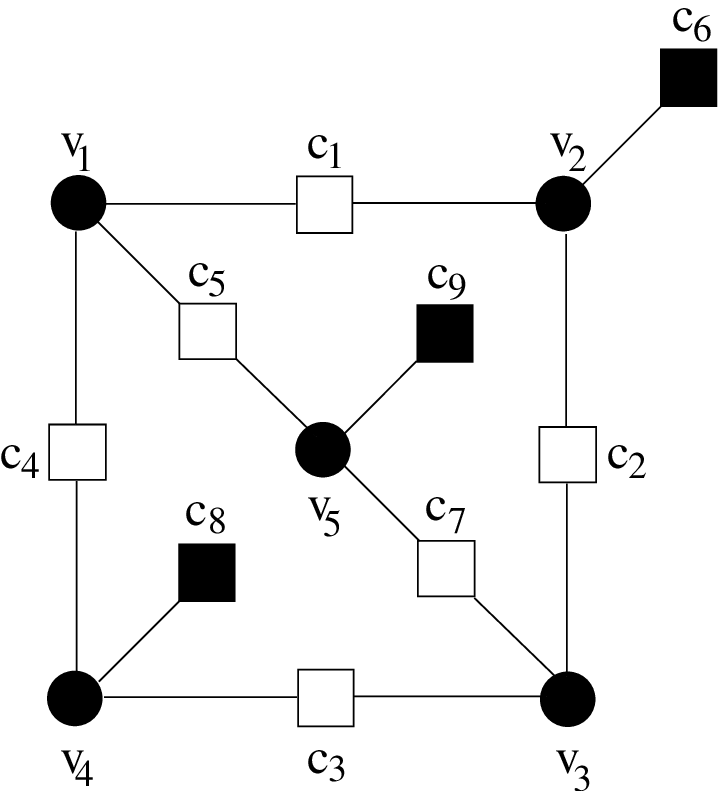}
}
\caption{Subgraphs isomorphic to \subref{44trappingset} a $(4,4)$ trapping set \subref{53trappingset} a $(5,3)$ trapping set} \label{4trappingsets}
\end{figure*}

\textit{Proof of Lemma \ref{lemma4}:} Let $\mathcal{T}_1=\{v_1, v_2,v_3,v_4\}$ be the variable nodes that form an eight cycle (see Fig.\ref{44trappingset}) . If no two checks from $\{c_5,c_6,c_7,c_8\}$ are connected to a common variable node, then $\mathcal{T}_1$ is a $(4,4)$ trapping set  and hence a failure set of size four. On the other hand, if $\mathcal{T}_1$ is not a trapping set, then there must be at least one variable node which is connected to two checks from $\{c_5,c_6,c_7,c_8\}$. Assume that $c_5$ and $c_7$ are connected to $v_5$ and the third check of $v_5$ is $c_9$ (see Fig.\ref{53trappingset}). We claim that $\mathcal{T}_2=\mathcal{T}_1\cup \{v_5\}$ is a failure set. Let $\mathcal{E}$ and $\mathcal{O}$ be as defined in Theorem \ref{thm1}.

\textbf{Case 1:} No two checks from $\mathcal{O}=\{c_6,c_8,c_9\}$ are connected to a common variable node. Then $\mathcal{T}_2$ is a $(5,3)$ trapping set and hence a failure set of size five.

\textbf{Case 2:} All the three checks in $\mathcal{O}$ are connected to a common variable node, say $v_6$. Then $\mathcal{T}_2 \cup \{v_6\}$ is a codeword of weight six and it is easy to see that $\mathcal{T}_2$ is a failure set. 

\textbf{Case 3:} There are variable nodes connected to two checks from $\mathcal{O}$. There can be at most two such variable nodes (if there are three such variable nodes, they will form a cycle of length less than or equal to six violating the condition that the graph has girth eight). Note that if $S(\mathbf{r})=\mathcal{T}_2$, the decoder has a chance of correcting only if a check node in $\mathcal{E}$ receives an incorrect message from a variable node outside $\mathcal{T}_2$ in some $j^{th}$ iteration. We now prove that this is not possible. Indeed in the first iteration

\begin{eqnarray}
\omega_1(v,:)&=&\left\{\begin{array}{cl}
1, & v \in \mathcal{T}_2 \\
0, & \mbox{otherwise}
\end{array} \right. \nonumber 
\end{eqnarray}
By similar arguments as in the proof for Theorem \ref{thm1}, it can be seen that the only check nodes which send incorrect messages to variable nodes outside $\mathcal{T}_2$ are $c_6,c_8$ and $c_9$. There are now two subcases.

\textbf{Subcase 1:} There is one variable node connected to two checks from $\mathcal{O}$. Let $v_6$ be connected to $c_6$ and $c_8$. It can be seen that the third check connected to $v_6$ cannot belong to $\mathcal{E}$ as this would violate the girth condition. So, let the third check be $c_{10}$. In the first half of second iteration, we have
\begin{eqnarray}
\omega_2(v,c)&=&\left\{\begin{array}{cl}
1, & v \in \mathcal{T}_2 \mbox{~or~} (v,c)=(v_6,c_{10}) \\
0, & \mbox{otherwise}
\end{array} \right. \nonumber 
\end{eqnarray}
The only check nodes which send incorrect messages to variable nodes outside $\mathcal{T}_2$, are $c_6,c_8,c_9$ and $c_{10}$.  The variable node $v_6$ is connected to $c_6$ and $c_8$. If $c_9$ and $c_{10}$ are not connected to any common variable node, we are done. On the other hand, let $c_9$ and $c_{10}$ be connected to a variable node, say $v_7$. The third check of $v_7$ cannot be in $\mathcal{E}$. Proceeding as in the case of proof for Lemma \ref{lemma3}, we can prove that $\mathcal{T}_2$ is a failure set by observing that there cannot be a variable node outside $\mathcal{T}_2$ which sends an incorrect message to a check in $\mathcal{E}$.

\textbf{Subcase 2:} There are two variable nodes connected to two checks from $\mathcal{O}$. Let $c_6$ and $c_8$ be connected to $v_6$ and $c_6$ and $c_9$ connected to $v_7$. Proceeding as above, we can conclude that $\mathcal{T}_2$ is a failure set.
\section*{Acknowledgment}
This work is funded by NSF under Grant CCF-0634969, ITR-0325979 and INSIC-EHDR program. The authors would like to thank Anantharaman Krishnan for illustrations.

\end{document}